\documentclass[aps,preprint,groupedaddress,floatfix]{revtex4}
\usepackage{graphicx}
\begin{document}

\title{ Exotics at Our Home}
\author {
Nikolay Achasov }

\affiliation{ Laboratory of Theoretical Physics,
 Sobolev Institute for Mathematics, 630090, Novosibirsk, Russia
}

%\date{\today}

\begin{abstract}
\begin{enumerate}
\item Light Scalars as  Four-Quark States
\item Isotensor Tensor $E(1500-1600)$ State
\item $X(3872)$ State as  Charmonium $\chi_{c1}(2P)$
\item Two-gluon Annihilation of Charmonium $\chi_{c2}(2P)$
\end{enumerate}
\end{abstract}

\maketitle
\section{Introduction}
\label{intro} I would like to discuss the exotic phenomena  that I
know not by hearsay. That is, the exotic phenomena, which we
analyzed, explained and made the new predictions, the considerable
part of which is confirmed already by experiment, and other ones
wait the experimental checking.
\section{Light Scalars as  Four-quark States }
\label{sec-1} The $a_0(980)$ and $f_0(980)$ mesons are
well-established parts of the proposed light scalar meson nonet
\cite{pdg-2018}. From the beginning, the $a_0(980)$ and $f_0(980)$
mesons became one of the central problems of nonperturbative QCD,
as they are important for understanding the way chiral symmetry is
realized in the low-energy region and, consequently, for
understanding confinement. Many experimental and theoretical
papers have been devoted to this subject.

There is much evidence that supports the four-quark model of light
scalar mesons \cite{jaffe,weinberg}.

The suppression of the $a^0_0(980)$ and $f_0(980)$ resonances in
the $\gamma\gamma\to\eta\pi^0$ and $\gamma\gamma\to\pi\pi$
reactions, respectively, was predicted in 1982 \cite{fourQuarkGG},
$\Gamma_{a^0_0\gamma\gamma}\approx \Gamma_{f_0\gamma\gamma}\approx
0.27$ keV, and confirmed by experiment \cite{pdg-2018}. The high
quality Belle data  \cite{uehara2008,uehara2009} allowed to
elucidate the mechanisms of the $\sigma(600)$, $f_0(980)$, and
$a^0_0(980)$ resonance production in $\gamma\gamma$ collisions
confirmed their four-quark structure \cite{AS88,annsgnGamGam}.
Light scalar mesons are produced in $\gamma\gamma$ collisions
mainly via rescatterings, that is, via the four-quark transitions.
As for $a_2(1320)$ and $f_2(1270)$ (the well-known $q\bar q$
states), they are produced mainly via the two-quark transitions
(direct couplings with $\gamma\gamma$). As a result the
practically model-independent prediction of the  $q\bar q$ model
$g^2_{f_2\gamma\gamma}:g^2_{a_2\gamma\gamma}=25:9$ agrees with
experiment rather well. As to the ideal  $q\bar q$ model
prediction $g^2_{f_0\gamma\gamma}:g^2_{a_0\gamma\gamma}=25:9$, it
is excluded by experiment.

The argument in favor of the four-quark nature of $a_0(980)$ and
$f_0(980)$ is the fact that the $\phi(1020)\to a^0_0\gamma$ and
$\phi(1020)\to f_0\gamma$ decays go through the kaon loop:
$\phi\to K^+K^-\to a^0_0\gamma$, $\phi\to K^+K^-\to f_0\gamma$,
i.e., via the four-quark transition
\cite{achasov-89,achasov-97,a0f0,our_a0,achasov-03}. The kaon-loop
model was suggested in Ref. \cite{achasov-89} and confirmed by
experiment ten years later \cite{SNDa0,f0exp,kloea0}.

It was shown in Ref. \cite{achasov-03} that the production of
$a^0_0(980)$ and $f_0(980)$ in $\phi\to a^0_0\gamma\to
\eta\pi^0\gamma$ and $\phi\to f_0\gamma\to\pi^0\pi^0\gamma$ decays
is caused by the four-quark transitions, resulting in strong
restrictions on the large-$N_C$ expansion  of the decay
amplitudes. The analysis showed that these constraints give new
evidence in favor of the four-quark nature of the $a_0(980)$ and
$f_0(980)$ mesons.

In Refs. \cite{agsh,ak-07} it was shown that the description of
the $\phi\to K^+K^-\to\gamma a^0_0(980)/f_0(980)$ decays requires
virtual momenta of $K (\bar K)$ greater than $2$ GeV, while in the
case of loose molecules with a binding energy about 20 MeV, they
would have to be about 100 MeV. Besides, it should be noted that
the production of scalar mesons in the pion-nucleon collisions
with large momentum transfers also points to their compactness
\cite{ADS98}.

It was also shown in Refs. \cite{annshgn-94,annshgn-07} that the
linear $S_L(2)\times S_R(2)$ $\sigma$ model \cite{gellman}
reflects all of the main features of low-energy $\pi\pi\to\pi\pi$
and $\gamma\gamma\to\pi\pi$ reactions up to energy 0.8 GeV and
agrees with the four-quark nature of the $\sigma$ meson. This
allowed for the development of a phenomenological model with the
right analytical properties in the complex $s$ plane that took
into account the linear $\sigma$ model, the $\sigma(600)-f_0(980)$
mixing and the background \cite{our_f0_2011}. This background has
a left cut inspired by crossing symmetry, and the resulting
amplitude agrees with results obtained using the chiral expansion,
dispersion relations, and the Roy equation \cite{sigmaPole}, and
with the four-quark nature of the $\sigma(600)$ and $f_0(980)$
mesons as well. This model well describes the experimental data on
$\pi\pi\to\pi\pi$ scattering up to $1.2$ GeV.

Moreover, the absence of $J/\psi\to \gamma f_0(980),\ \rho
a_0(980),\ \omega f_0(980)$ decays in the presence of intense
$J/\psi\to \gamma f_2(1270),\ \gamma f'_2(1525),\ \rho a_2(1320),\
\omega f_2(1270)$ decays is at variance with the $P$-wave
two-quark, $q\bar q$, structure of $a_0(980)$ and $f_0(980)$
resonances \cite{achasov-1998}.

It is shown in Ref. \cite{correlation} that the recent data on the
$K^0_S K^+$ correlation in Pb-Pb interactions Ref.
\cite{alice-2017} agree with the data on the
$\gamma\gamma\to\eta\pi^0$ and $\phi\to\eta\pi^0\gamma$ reactions
and support the four-quark model of the $a_0(980)$ meson. It is
shown that the data does not contradict the validity of the
Gaussian assumption.

In Refs. \cite{dsdecay,dsdecayConf} it was suggested the program
of studying light scalars in semileptonic $D$ and $B$ decays,
which are the unique probe of the $q\bar q$ constituent pair in
the light scalars. We studied production of scalars $\sigma(600)$
and $f_0(980)$ in the $D_s^+\to s\bar s\, e^+\nu\to\pi^+\pi^-\,
e^+\nu$ decays, the conclusion was that the fraction of the $s\bar
s$ constituent components in $\sigma(600)$ and $f_0(980)$ is
small. Unfortunately, at the moment the CLEO statistics
 \cite{cleo} is rather poor, and thus new high-statistics data
are highly desirable.

It was noted in Refs. \cite{dsdecay,dsdecayConf} that no less
interesting is the study of semileptonic decays of $D^0$ and $D^+$
mesons -- $D^+\to d\bar d\, e^+\nu\to
[\sigma(600)+f_0(980)]e^+\nu\to \pi^+\pi^-e^+\nu$, $D^0\to d\bar
u\, e^+\nu\to a_0^-e^+\nu\to\pi^-\eta e^+\nu$ and $D^+\to d\bar
d\, e^+\nu\to a_0^0 e^+\nu\to\pi^0\eta e^+\nu$ (or the
charged-conjugated ones) which had not been investigated. It is
very tempting to study light scalar mesons in semileptonic decays
of $B$ mesons \cite{dsdecayConf}: $B^0\to d\bar u\, e^+\nu\to
a_0^-e^+\nu\to\pi^-\eta e^+\nu$, $B^+\to u\bar u\, e^+\nu\to a_0^0
e^+\nu\to\pi^0\eta e^+\nu$, $B^+\to u\bar u\, e^+\nu\to
[\sigma(600)+f_0(980)]e^+\nu\to \pi^+\pi^-e^+\nu$.

\begin{figure}[h]
\centering
\includegraphics[width=10cm,clip]{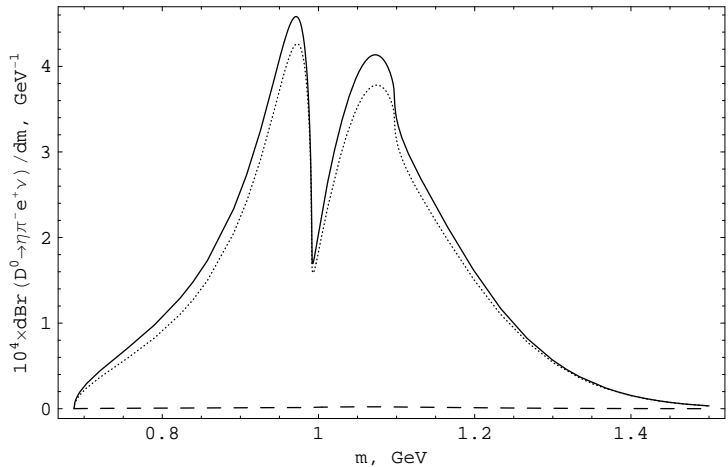}
\caption{The plot of $D^0\to (a^-_0,a'^-_0)\, e^+\nu\to\eta\pi^-
e^+\nu$ spectrum without a coupling of $a_0^-(980)$ with the
constituent $d\,\bar{u} $ state. The solid line is the total
contribution, the dotted line is caused by
 the mixing $a_0^{\prime-}(1400)\to
a_0^-(980)$, the dashed line is  caused by the $a_0'^-(1400)$
production, Ref. \cite{NNA+AVK}.} \label{fig-1}
\end{figure}

Recently BESIII Collaboration measured the decays $D^0\to d\bar
u\, e^+\nu\to a_0^-e^+\nu\to\pi^-\eta e^+\nu$ and $D^+\to d\bar
d\, e^+\nu\to a_0^0 e^+\nu\to\pi^0\eta e^+\nu$ for the first time
\cite{besIII}. In Ref. \cite{NNA+AVK} we discuss the Ref.
\cite{dsdecay} program in light of these measurements taking into
account contribution of $a_0'$ meson with mass about $1400$ MeV. A
variant when $a_0^-(980)$ has no $q\bar q$ constituent component
at all is presented in Figure \ref{fig-1}, that is, $a_0^-(980)$
is produced as a result of mixing $a_0^{\prime-}(1400)\to
a_0^-(980)$,
 $D^0\to d\bar u\, e^+\nu\to a_0^{\prime -}e^+\nu \to a_0^-e^+\nu\to \pi^-\eta e^+\nu$.

The first measurement of BESIII is the important step for the
investigation of light scalar mesons nature, but for the present
the statistics is not adequate for the conclusions.

\section{Isotensor Tensor \bf\boldmath{ $E(1500-1600)$} State}
\label{sec-2}
 Thirty six years ago we predicted \cite{fourQuarkGG} the striking
interference picture in the  $\gamma\gamma\to\rho^0\rho^0$ and
$\gamma\gamma\to\rho^+\rho^-$ reactions  in the $q^2\bar q^2$ MIT
model \cite{jaffe}.

\begin{figure}[h]
\centering
\includegraphics[width=10cm,clip]{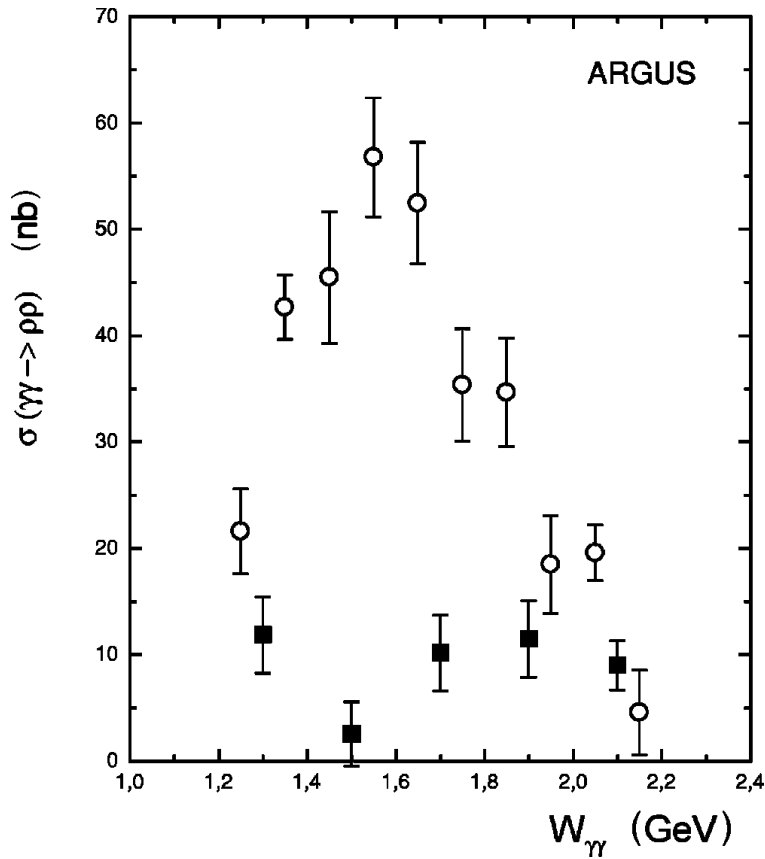}
\caption{The white circles are  $\gamma\gamma\to\rho^0\rho^0$; the
black small squares are  $\gamma\gamma\to\rho^+\rho^-$.}
\label{fig-2}
\end{figure}

We explained the strong boost near the threshold in the
$\gamma\gamma\to\rho^0\rho^0$ reaction by the production of the
isotensor tensor and isoscalar tensor resonances,  then the
destructive interference of their contributions in the
$\gamma\gamma\to\rho^+\rho^-$ reaction follows from isotopic
symmetry!\\[3pt]
 Experiment backed up this prediction, JADE 1983, ARGUS 1991,
 see Figure \ref{fig-2} and Refs. \cite{I=21985, I=21991}.

We believe that the Belle data will support the above picture.

 We
\cite{I=21999} hope that JEFLAB will find the charged components
of the isotensor tensor state $E^{\pm}$ in the mass spectra of the
$\rho^{\pm}\rho^0$ states in the reactions $\gamma
N\to\rho^{\pm}\rho^0 N(\Delta)$.

We \cite{I=21991,I=21992} believe also that IHEP in Protvino could
find $E^{\pm\pm}$ in the mass spectra  of the
$\rho^{\pm}\rho^{\pm}$ states in the reactions $\pi
N\to\pi\rho^{\pm}\rho^{\pm}N(\Delta)$ and $N N\to N(\Delta)
\rho^{\pm}\rho^{\pm}N(\Delta)$.

\section{{\bf\boldmath{$X(3872)$ State as  Charmonium $\chi_{c1}(2P)$}, Refs.  \cite{NNA+EVR,NNA+EVR+,NNA3872}}}
\label{sec-3}

The two dramatic discoveries have generated a stream of the
$D^{*0}\bar D^0+D^0\bar D^{*0}$  molecular interpretations of the
$X(3872)$ resonance.
 The mass of the $X(3872)$ resonance is 50 MeV lower than predictions of the most lucky naive  potential
models for the mass of the $\chi_{c1}(2P)$ resonance,
\begin{equation}
\label{shiftmass} m_X-m_{\chi_{c1}(2P)}= -\Delta\approx -
50\,\mbox{MeV},\nonumber
\end{equation}
and the relation between the branching ratios
\begin{equation}
\label{isotopicviolation}
 BR(X\to\pi^+\pi^-\pi^0J/\psi(1S))\sim
BR(X\to\pi^+\pi^-J/\psi(1S))\,\nonumber,
\end{equation}
that is interpreted as a strong violation of isotopic symmetry.

But, the bounding energy is small, $\epsilon_B\lesssim 1$ MeV.
That is, the radius of the molecule is large, $r_{X(3872)}\gtrsim
5 = 5\cdot 10^{-13}$ cm. As for the charmonium, its radius is less
one  fermi, $r_{\chi_{c1}(2P)}\lesssim$  fermi $= 10^{-13}$ cm.

 That is, the molecule volume is  $125\div 1000$ times
as large as the charmonium volume,
$V_{X(3872)}/V_{\chi_{c1}(2P)}\gtrsim$  $125\div 1000$. This means
a probability of production of a giant molecule in hard processes,
at small distances, is suppressed in comparison with a probability
of production of heavy a charmonium by  a factor  $\sim
V_{\chi_{c1}(2P)}/V_{X(3872)}$.

But, in reality
\cite{LHCb2012,NNA+EVR+,NNA3872,pdg-2018,Belle2009}
\begin{equation}
 0.74<\frac{\sigma (pp \to X(3872)
+ anything)}{ \sigma (pp\to\psi(2S) + anything)} <2.1.\nonumber
\end{equation}
 with rapidity in
the range 2,5 - 4,5 and transverse momentum in the range 5-20 GeV.

  In addition, \cite{Belle2015,NNA3872,pdg-2018,Belle2009}
\begin{equation}
 0.2<\frac{BR(B^0\to X(3872)K^+\pi^-)}{BR(B^0\to\psi(2S)K^+\pi^-)}<0.6.\nonumber
\end{equation}
The extended molecule is produced in hard processes as intensively
as the compact charmonium.  It's miracle!

We explain the shift of  the mass of the
 $X(3872)$ resonance with respect to the prediction of a potential model for the mass of the
$\chi_{c1}(2P)$ charmonium by the contribution of the virtual
$D^*\bar D+c.c.$ intermediate states into the self energy of the
$X(3872)$ resonance \cite{NNA+EVR+,NNA3872}.

 This allows us to estimate the coupling constant of the
$X(7872)$ resonance with the $D^{*0}\bar D^0$ channel, the
branching ratio of the $X(3872) \to D^{*0}\bar D^0 + c.c.$ decay,
and the branching ratio of  the $X(3872)$ decay into all
non-$D^{*0}\bar D^0 + c.c.$ states
\cite{NNA+EVR,NNA+EVR+,NNA3872}.

 We \cite{NNA+EVR+,NNA3872} predict that the hadron channels of the decays of
$\chi_{c1}(2P)$ via two gluon ( $X(3872)\to gluon\ gluon\to
hadrons$) should be the same as in the  $\chi_{c1}(1P)$ case, that
is, there should be a few tens of such channels. The discovery of
these decays would be the strong (if not decisive) confirmation of
our scenario.

 As for $BR(X\to\omega J/\psi)\sim BR(X\to\rho
J/\psi)$, this could be a result of dynamics. In our scenario the
$\omega J/\psi$ state is produced via the three gluons.

 As for the $\rho J/\psi$ state, it
is produced both via the one photon, and via the three gluons (via
the contribution $\sim m_u-m_d$ ).

 Close to our scenario is an example of the
$J/\psi\to\rho\eta'$ and $J/\psi\to\omega\eta'$  decays. According
to Ref. \cite{pdg-2018}
\begin{equation}
\label{Jpsi}
 BR(J/\psi\to\rho\eta')=(1.05\pm 0.18)\cdot10^{-4}\ \ \mbox{and}\ \
BR(J/\psi\to\omega\eta')=(1.82\pm 0.21)\cdot10^{-4}.
\end{equation}
Note that in the $X(3872)$ case the $\omega$ meson is produced on
its tail ($m_X-m_{J/\psi}=775$ MeV), while the $\rho$ meson is
produced on a half.

It is well known that the physics of charmonium ($c\bar c$) and
bottomonium ($b\bar b$) is similar. Let us compare the already
known features of X(3872) with the ones of $\Upsilon_{b1}(2P)$.

 The LHCb Collaboration published a landmark result
\cite{LHCb2014}
\begin{equation}
\label{Xtogamma}
 \frac{BR(X\to\gamma \psi(2S))}{BR(X\to\gamma
J/\psi)}=C_X\left(\frac{\omega_{\psi(2S)}}{\omega_{J/\psi}}\right)^3=2.46\pm
0.7\,,
\end{equation}
where $\omega_{\psi(2S)}$ and $\omega_{J/\psi}$ are the energies
of the photons in the $X\to\gamma \psi(2S)$ and $BR(X\to\gamma
J/\psi)$ decays, respectively.

 On the other hand, it is known \cite{pdg-2018} that
 \begin{equation}
 \label{chib1(2P)}
\hspace*{-18pt}\frac{BR(\chi_{b1}(2P)\to\gamma
\Upsilon(2S))}{BR(\chi_{b1}(2P)\to\gamma
\Upsilon(1S))}=C_{\chi_{b1}(2P)}\left(\frac{\omega_{\Upsilon(2S)}}{\omega_{\Upsilon(1S)}}\right)^3=2.16\pm0.28\,,
\end{equation}
where $\omega_{\Upsilon(2S)}$ and $\omega_{\Upsilon(1S)}$ are the
energies of the photons in the $\chi_{b1}(2P)\to\gamma
\Upsilon(2S)$ and $\chi_{b1}(2P)\to\gamma \Upsilon(1S)$ decays,
respectively.

Consequently,
\begin{equation}
\label{CX}
 C_X=136.78\pm38.89
\end{equation}
and
\begin{equation}
\label{Cchib1(2P)}
 C_{\chi_{b1}(2P)}=80\pm10.37\
\end{equation}
as all most lucky versions of the  potential model predict for the
quarkonia, $C_{\chi_{c1}(2P)}\gg 1$ and $C_{\chi_{b1}(2P)}\gg 1$.

 According to Ref. \cite{pdg-2018}
\begin{equation}
\label{omegaUpsilon(1S)}
 BR(\chi_{b1}(2P)\to\omega\Upsilon(1S))=\left
(1.63\pm^{0.4}_{0.34}\right )\%\,.
\end{equation}

If  the one-photon mechanism dominates in the $X(3872)\to\rho
J/\psi$ decay then one should expect
\begin{equation}
\label{onephotonrhoUpsilon(1S)}
 BR(\chi_{b1}(2P)\to\rho\Upsilon(1S))\sim(e_b/e_c)^2\cdot 1.6\,\%=(1/4)\cdot
1.6\,\%= 0.4\%\,,
\end{equation}
 where $e_c$ and
$e_b$ are the charges of the $c$ and $b$ quarks, respectively.

If  the three-gluon mechanism (its part $\sim m_u-m_d$ ) dominates
in the $X(3872)\to\rho J/\psi$ decay then one should expect
\begin{equation}
\label{threegluonrhoUpsilon(1S)}
 BR(\chi_{b1}(2P)\to\rho\Upsilon(1S))\sim 1.6\%\,.
\end{equation}

We believe that  discovery of a significant number of unknown
decays of $X(3872)$ into non-$D^{*0}\bar D^0 +c.c.$ states via two
gluons and discovery of the $\chi_{b1}(2P)\to\rho\Upsilon(1S)$
decay could decide destiny of $X(3872)$.

Once more, we discuss the scenario where the $\chi_{c1}(2P)$
charmonium sits on the $D^{*0}\bar D^0$ threshold but not a mixing
of the giant $D^{*}\bar D$ molecule and the compact
$\chi_{c1}(2P)$ charmonium. Note that the mixing of such states
requests the special justification. That is, it is necessary to
show that the transition of the giant molecule into the compact
charmonium is considerable at insignificant overlapping of their
wave functions. Such a transition
$\sim\sqrt{V_{\chi_{c1}(2P)}/V_{X(3872)}}$  and a branching ratio
of a decay via such a transition $\sim
V_{\chi_{c1}(2P)}/V_{X(3872)}$.

Note that now the $X(3872)$ state is named in Ref. \cite{pdg-2018}
as $\chi_{c1}(3872)$.

 The above scenario can be checked in the process
$e^+e^-\to\psi(4040)\to\gamma\,(gluon\
gluon)\to\gamma\chi_{c1}(3872)\to\gamma\,(gluon\
gluon)\to\gamma\,( light\ hadrons)$ on BES III, for example, or on
the super c-tau factory that is projecting in Novosibirsk.

\section{Two-gluon Annihilation of Charmonium \bf\boldmath{$\chi_{c2}(2P)$}}
\label{sec-4}

 We \cite{NNA+XWK} expect that $BR(\chi_{c2}(2P)\to gluon\,
gluon)\gtrsim 2 \%$  if the Particle Data Group as well as the
BaBar and Belle collaborations have correctly identified the
state.

 In reality, this branching ratio
corresponds to the one for $\chi_{c2}(2P)$ decaying into light
hadrons. The hadron channels of the two-gluon decays of
$\chi_{c2}(2P)$ should be the same as in the $\chi_{c2}(1P)$ case,
that is, there should be a few tens of  such channels.

 The ratio
of the two-photon and two-gluon widths of the charmonium decays
does not depend on the wave function in the  nonrelativistic
potential model of charmonium \cite{iteph}. It allows to find the
low limit of $BR(\chi_{c2}(2P)\to gluon\, gluon)$. The comparison
with the well-known data about  $\chi_{c2}(1P)$ allows us to
conclude that $BR(\chi_{c2}(2P)\to2 g)\approx (6.5\pm 2.0)\%$ is
very likely.

 The confirmation of the $\chi_{c2}(2P)$ state can be
tested by BESIII, for example, through the process
$e^+e^-\to\psi(4040)\to \gamma \chi_{c2}(2P)$. The search for
two-gluon decays of the $\chi_{c2}(2P)$ state is feasible for
BESIII as well as other super factories such as BaBar and Belle.

Note that now the $\chi_{c2}(2P)$ state is named in Ref.
\cite{pdg-2018} as $\chi_{c2}(3930)$.

\section{Acknowledgments}

  I am grateful to
Organizers of QUARKS-2018 for the kind Invitation.

The work was supported by the program of fundamental scientific
researches of the SB RAS No. II.15.1., project No. 0314-2016-0021
and  partially  by the Russian Foundation for Basic Research Grant
No. 16-02-00065.

%\newpage


\begin{thebibliography}{99}
\bibitem{pdg-2018}
M. Tanabashi et al. (Particle Data Group), Phys. Rev. D {\bf 98},
030001 (2018).
\bibitem{jaffe}
R.L. Jaffe, Phys. Rev. D {\bf 15}, 267 (1977);\\ Phys. Rev. D {\bf
15}, 281 (1977).

\bibitem{weinberg}
S. Weinberg, Phys. Rev. Lett. {\bf 110}, 261601 (2013).

\bibitem{fourQuarkGG}
N.N. Achasov, S.A. Devyanin, and G.N. Shestakov, Phys. Lett. B
{\bf 108}, 134 (1982);\\ Z. Phys. C {\bf 16}, 55 (1982).

\bibitem{uehara2008}
S. Uehara {\it et al.} (Belle Collaboration), Phys. Rev. D {\bf
78}, 052004 (2008).

\bibitem{uehara2009}
S. Uehara {\it et al.} (Belle Collaboration), Phys. Rev. D {\bf
80}, 032001 (2009).

\bibitem{AS88}
N.N. Achasov and G.N. Shestakov, Z. Phys. C {\bf 41}, 309 (1988).

\bibitem{annsgnGamGam}
N.N. Achasov and G.N. Shestakov, Phys. Rev. D {\bf 77}, 074020
(2008);\\ Phys. Rev. D {\bf 81}, 094029 (2010);\\ Usp. Fiz. Nauk
{\bf 54}, 799 (2011) [Sov. Phys. Usp. {\bf 181}, 827 (2011)].

\bibitem{achasov-89}
N.N. Achasov and V.N. Ivanchenko, Nucl. Phys. B {\bf 315}, 465
(1989).

\bibitem{achasov-03}
N.N. Achasov, Nucl. Phys. A {\bf 728}, 425 (2003).

\bibitem{a0f0}
N.N. Achasov and V.V. Gubin, Phys. Rev. D {\bf 63}, 094007
(2001);\\ N.N. Achasov and A.V. Kiselev, Phys. Rev. D {\bf 73},
054029 (2006).

\bibitem{our_a0}
N.N. Achasov and A.V. Kiselev, Phys. Rev. D {\bf 68}, 014006
(2003).

\bibitem{achasov-97}
N.N. Achasov and V.V. Gubin, Phys. Rev. D {\bf 56}, 4084 (1997).

\bibitem{SNDa0}
M.N. Achasov {\it et al.} (SND Collaboration), Phys. Lett. B {\bf
438}, 441 (1998);\\ M.N. Achasov {\it et al.}, Phys. Lett. B {\bf
479}, 53 (2000).

\bibitem{f0exp}
M.N. Achasov {\it et al.} (SND Collaboration), Phys. Lett. B {\bf
440}, 442 (1998);\\ M.N.Achasov {\it et al.}, Phys. Lett. B {\bf
485}, 349 (2000);\\ R.R. Akhmetshin {\it et al.} (CMD-2
Collaboration) Phys. Lett. B {\bf 462}, 380 (1999);\\ A.Aloisio
{\it et al.} (KLOE Collaboration) Phys. Lett. B {\bf 537}, 21
(2002);\\ C. Bini, P. Gauzzi, S. Giovanella, D. Leone, and S.
Miscetti, KLOE Note 173 06/02, http://www.lnf.infn.it/kloe/.

\bibitem{kloea0}
A.Aloisio {\it et al.} (KLOE Collaboration) Phys. Lett. B {\bf
536}, 209 (2002).

\bibitem{agsh}
N.N. Achasov, V.V. Gubin, and V.I. Shevchenko, Phys. Rev. D {\bf
56}, 203 (1997).

\bibitem{ak-07}
N.N. Achasov and A.V. Kiselev, Phys. Rev. D {\bf 76}, 077501
(2007);\\ Phys. Rev. D {\bf 78}, 058502 (2008).

\bibitem{ADS98}
N.N. Achasov and G.N. Shestakov, Phys. Rev. D {\bf 58}, 054011
(1998).

\bibitem{annshgn-94}
N.N. Achasov and G.N. Shestakov, Phys. Rev. D {\bf 49}, 5779
(1994).

\bibitem{annshgn-07}
N.N. Achasov and G.N. Shestakov, Phys. Rev. Lett. {\bf 99}, 072001
(2007).

\bibitem{gellman}
M. Gell-Mann and M. Levy, Nuovo Cimento {\bf 16}, 705 (1960).

\bibitem {our_f0_2011}
N.N. Achasov and A.V. Kiselev, Phys. Rev. D {\bf 83}, 054008
(2011);\\ Phys. Rev. D {\bf 85}, 094016 (2012).

\bibitem{sigmaPole}
I. Caprini, G. Colangelo and H. Leutwyler, Phys. Rev. Lett. {\bf
96}, 132001 (2006).

\bibitem{achasov-1998}
N.N. Achasov, Usp. Fiz. Nauk {\bf 41}, 1257 (1998) [Phys. Usp.
{\bf 41}, 1149 (1998)];\\ Yad. Fiz. {\bf 65}, 573 (2002) [Phys.
At. Nucl. {\bf 65}, 546 (2002)].

\bibitem{correlation}
N.N. Achasov and A.V. Kiselev, Phys. Rev. D {\bf 97}, 036015
(2018).

\bibitem{alice-2017}
S. Acharya {\it et al.} (ALICE Collaboration), Phys. Lett. B {\bf
774}, 64 (2017).

\bibitem{dsdecay}
N.N. Achasov and A.V. Kiselev, Phys. Rev. D {\bf 86}, 114010
(2012).

\bibitem{dsdecayConf}
N.N. Achasov and A.V. Kiselev, Int. J. Mod. Phys. Conf. Ser. {\bf
35}, 1460447 (2014),\\
http://www.worldscientific.com/doi/pdf/10.1142/S2010194514604475.

\bibitem{cleo}
K.M Ecklund et al. CLEO Collaboration, Phys. Rev. D {\bf 80},
052009 (2009).

\bibitem{besIII}
M. Ablikim {\it et al.} (BESIII Collaboration), arXiv:1803.02166.


\bibitem{NNA+AVK}
N.N. Achasov and A.V. Kiselev, arXiv:1805.10145.

\bibitem{I=21985}
 N.N. Achasov and G.N. Shestakov,  Z. Phys. {\bf C} 27, 99 (1985).

\bibitem{I=21991}
 N.N. Achasov and G.N. Shestakov,  Sov. Phys. Usp. {\bf 34} (6), 471  (1991).

\bibitem{I=21999}
 N.N. Achasov and G.N. Shestakov, Phys. Rev. D {\bf 60}, 114021
 (1999).

\bibitem{I=21992}
N.N. Achasov and G.N. Shestakov, Int. J. Mod. Phys. A {\bf 7} 4313
(1992).

\bibitem{NNA+EVR}
 N.N. Achasov and E.V. Rogozina, JETP Lett. {\bf 100}, 227 (2014).

\bibitem{NNA+EVR+}
N.N. Achasov and E.V. Rogozina, Mod. Phys. Lett. A {\bf 30},
1550181 (2015);\\
 J.Univ.Sci.Tech.China {\bf 46}, 574 (2016).

\bibitem{NNA3872}
 Nikolay Achasov, EPJ Web Conf. {\bf 125}, 04002 (2016);\\
N.N. Achasov, Phys. Part. Nucl. {\bf 48}, 839 (2017).

\bibitem{LHCb2012}
R. Aaij et al. (LHCb Collaboration), Eur. Phys.J. C {\bf 72}, 1972
(2012).

\bibitem{Belle2009} C.-Z. Yuan (Belle Collaboration),  \textit{ Proceedings of the XXIX PHYSICS IN COLLISION}
Kobe, Japan, arXiv: 0910.3138 [hep-ex] (2009)

\bibitem{LHCb2014}
R. Aaij et al. (LHCb Collaboration), Nucl. Phys. B {\bf 886}, 665
(2014).

\bibitem{Belle2015} A. Bala, et al. (Belle Callaboration), Phys.
Rev. D  {\bf 91} 051101(R) (2015)

\bibitem{NNA+XWK}
N.N. Achasov and Kang Xian-Wei, Chin. Phys. C {\bf 41}, 123102
(2017).

\bibitem{iteph}
V.A. Novikov, L.B. Okun, M.A. Shifman, A.I. Vainshtein, M.B.
Voloshin, and\\ V.I. Zakharov, Phys. Rept. {\bf 41}, 1 (1978).

\end{thebibliography}
\end{document}